\documentstyle[twocolumn,prl,aps,epsf]{revtex}
\begin{document}

\title{Supersonic and multiple topological excitations
in the driven Frenkel-Kontorova model with exponential interaction}

\author{O.\ M.\ Braun$^{*}$}

\address{
Institute of Physics,
National Ukrainian Academy of Sciences,
03650 Kiev-39, Ukraine}

\maketitle

\begin{abstract}

The criteria for the existence of supersonic and multiple
topological excitations (kinks)
in the driven Frenkel-Kontorova model
(a chain of atoms placed into
an external periodic potential)
with anharmonic (exponential)
interatomic interactions are studied.

\end{abstract}


\section{Introduction}
\label{intro}

Topological excitations
play a very important role in system dynamics,
because they are responsible for 
mass and charge transport in
solids and on crystal surfaces.
As classical examples, one can mention
dislocations and crowdions
described by the Frenkel-Kontorova (FK) model 
\cite{FK38},
where the topological excitations correspond
to kinks that describe local compression or
expansion of a commensurate structure. 
The FK model has numerous applications in
superionic conductivity,
surface physics,
hydrogen-bonded chains,
Josephson junctions,
tribology, etc.\ 
(e.g., see \cite{BKrev} and references therein).

In the continuum limit approximation,
the equation of motion
of the FK model reduces to 
the exactly integrable sine-Gordon (SG) equation.
But in continuum models,
even in a model with anharmonic (but local)
interaction, 
the topological excitations are always subsonic, 
the kink cannot propagate with a velocity $v$ 
larger than the sound speed $c$
because of Lorentz contraction of kink's width.
Moreover, in the classical FK model, the kinks 
of the same topological charge
repel from one another and, therefore,
they cannot carry a multiple topological charge. 

\begin{figure}
\epsfxsize=\hsize
\epsfbox{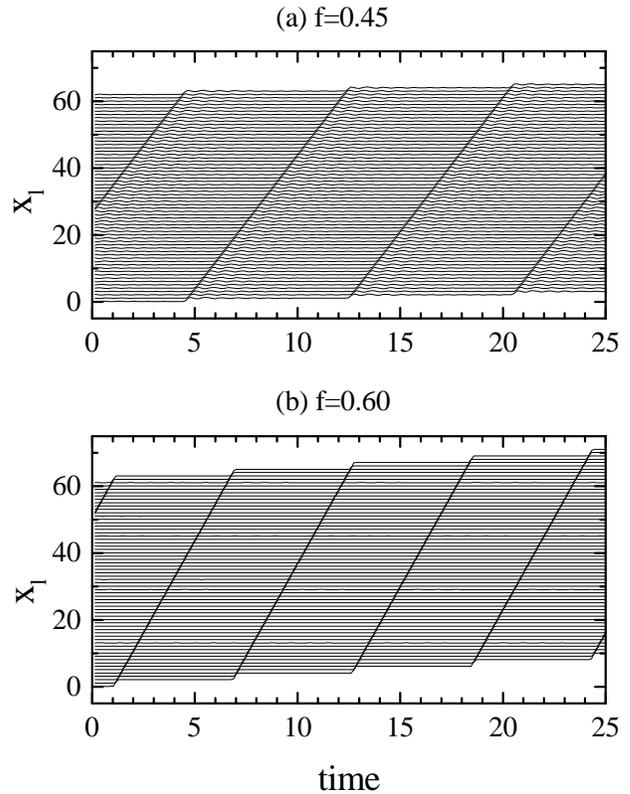}
\caption{
Atomic trajectories of the FK model
with exponential 
interaction
for $\beta=1/\pi$, $g=1$, and $\eta=0.05$.
(a) The single supersonic kink, 
$f=0.45$, $v_{\rm k}/c \approx 1.28$, and
(b) the double kink, 
$f=0.60$, $v_{\rm k}/c \approx 1.75$.}
\label{Fig01}
\end{figure}

However, simulation demonstrates that
supersonic kinks as well as multiple kinks do exist.
For example, Fig.\ \ref{Fig01}
shows the propagation of supersonic single and double kinks
in the FK model with exponential interatomic interaction.
The motion of topological solitons with 
{\it supersonic\/} velocities was firstly predicted, 
to the best of our knowledge, 
analytically by Kosevich and Kovalev \cite{KK73} 
in the FK model with some specific interatomic
interaction in the continuum limit approximation. 
Later, supersonic topological solitons were observed 
by Bishop {\it et al.\/} \cite{bishop84} 
in molecular dynamics study of polyacetylene. 
Then the supersonic kinks were studied
numerically in the discrete FK model 
with anharmonic interaction by Savin \cite{savin}. 
It was shown that for certain supersonic 
kink velocities, when its width coincides 
with that of the corresponding Toda soliton
\cite{Toda},
the kink propagates almost without energy losses. 
{\it Multiple\/} fast (but subsonic) kinks 
were firstly observed numerically 
by Peyrard and Kruskal \cite{PK84}
in the classical highly discrete FK model. 
Alfimov {\it et al.\/} \cite {AEKM93} 
have shown that multiple kinks exist also
in continuum systems with {\it nonlocal\/} interaction.
The bounded states of kinks 
(subsonic as well as supersonic) 
in the case of anharmonic interaction
were also studied numerically 
by Zolotaryuk {\it et al.\/} \cite{jaro}.
It was found that these multiple kinks are
asymptotically unstable.
The dynamics of the generalized FK chain
{\it driven\/} by a dc external force was studied numerically
in \cite{BBR97,traffic}, where we observed
the existence of supersonic kinks
and multiple (double and triple at least) kinks.
Recently, the existence of multiple kinks 
for certain kink velocities in the {\it discrete\/}
FK-type model was proven rigorously
\cite{Yu}.

The aim of the present paper is to find
the criteria for the existence of supersonic and
multiple topological excitations
in the FK-type models.
We will show that, 
first, the model must be {\it discrete\/}
as was already mentioned above.
Second, because kink's motion in a discrete chain
is damped due to radiation of phonons,
we must apply {\it a driving force\/} to support the motion
(this point was lost in the previous studies
\cite{KK73,bishop84,savin,PK84,AEKM93,jaro}).
Third, the interatomic interaction
must be {\it anharmonic\/}.
Under these three conditions, 
the model admits both 
{\it supersonic kinks\/} and {\it multiple kinks\/}.

The paper is organized as follows.
The model is introduced in Sec.\ \ref{model}.
Then in Sec.\ \ref{continuum} the problem is studied
with the help of the continuum limit approximation.
We show that if the discreteness effects are
properly taken into account,
the model formally allows both supersonic kinks
and multiple kinks.
Moreover, we prove that for 
a given set of parameters
the model admits either the single-kink solution
or the double-kink solution,
but not both solutions simultaneously.
Numerical solution of the corresponding ordinary
differential equation, which can be obtained
with any desired accuracy, does show
the existence of supersonic and multiple kinks.
In Secs.\ \ref{supersonic} and~\ref{multiple}
we develop an approximate variational approach
that helps to find the conditions
for existence of supersonic and multiple kinks.
We show that supersonic kinks may exist
in the model with anharmonic interaction only,
and the multiple kinks may be stable for
supersonic velocities only.
These conclusions are confirmed by simulation.
Finally, Sec.\ \ref{conclusion}
concludes the paper with the comment that
supersonic kinks and multiple kinks may be considered
as ``disturbed'' Toda solitons.

\section{Model}
\label{model}

We consider  the chain of atoms
placed into the sinusoidal substrate potential
and driven by a dc force $f$ applied to all atoms,
so that the equation of motion is
\begin{eqnarray}
\ddot x_l(t)
+\eta \dot x_l(t) 
&&
+\frac{\partial}{\partial x_l}
\left[ V(x_{l+1}-x_l) + V(x_l-x_{l-1}) \right] 
\nonumber \\
&&
+\sin x_l(t)-f=0, 
\label{mot}
\end{eqnarray}
where $x_l$ is the coordinate of the $l$th atom,
$\eta$ is the external viscous damping
coefficient introduced to compensate the driving force,
and
\begin{equation}
\label{pot}
V(x) = V_0 \exp (-\beta x) 
\end{equation}
describes the exponential repulsion
between nearest neighboring atoms.
We assume periodic boundary conditions 
with the number of atoms $N=N_s+p$, 
where $N_s$ is the number of wells of the periodic
potential, so we have one multiple $p$-kink 
($p$ excessive atoms)
inserted into the commensurate structure.
Similarly to the Toda model \cite{Toda},
one may add also an attractive linear interaction,
so that the interaction reduces to the harmonic one
in the limit $\beta \rightarrow 0$
(the classical FK model)
and to the hard--core interaction in the opposite
limit $\beta \rightarrow \infty$.
Throughout the paper, we use the dimensionless system
of units, where the atomic mass is $m=1$, 
the period of the external potential is $a=2\pi $ 
and its amplitude is $\varepsilon =2$. 
In these units, a characteristic
frequency of atomic vibration at a minimum 
of the external potential is $\omega _0=1$, 
a characteristic time scale is $\tau _0=2\pi $, 
and the maximum value of the external dc force,
when the minima of the sinusoidal
substrate potential disappear and 
the topological excitations cannot exist
anymore, is $f=1$. 

\section{Continuum approximation}
\label{continuum}

The systematic procedure to derive the equation of motion
in the continuum limit starting
from the discrete lattice 
was proposed by Rosenau \cite{Rosenau}.
For the anharmonic FK
model (\ref{mot}) it leads,
in the first order of the discreteness parameter,
to the equation
\begin{eqnarray}
\ddot{u}
&&
+\eta \dot{u}
-d^2 u^{\prime \prime }
\left(
1-\alpha d u^{\prime }
\right)
+\sin u - f
\nonumber \\
-
&&
h^2 \left[
\ddot{u}^{\prime \prime }
+\left( u^{\prime } \right)^2 \sin u
-u^{\prime \prime } \cos u \right] =0, 
\label{h18} 
\end{eqnarray} 
where 
$d=a \sqrt{g}$ is the width of the static kink, 
$g$ is the elastic constant defined as 
$
g=V^{\prime \prime }(a) 
=V_0 \beta^2 \exp (-\beta a), 
$
the anharmonicity
parameter $\alpha$ is defined as 
\begin{equation}
\label{a}
\alpha = -\frac{a}{d} \frac 
{V^{\prime \prime \prime }(a)} 
{V^{\prime \prime }(a)} 
=\frac{\beta}{\sqrt{g}},
\end{equation}
and the parameter
$h^2 = a^2 /12= \pi^2 /3$ describes
the discreteness effects.
Looking for a traveling--wave solution 
of Eq.\ (\ref{h18}) in the form 
$u(x,t)=u_{{\rm k}} (x-vt)
\equiv u_{{\rm k}} (z)$,
we obtain the equation
\begin{eqnarray}
h^2 v^2 u_{{\rm k}}^{
\prime \prime \prime \prime }
+
&&
\left( c^2-v^2-h^2 \cos u_{\rm k} \right)
u_{\rm k}^{\prime \prime }
+ h^2 \left( u_{\rm k}^{\prime }
\right)^2 \sin u_{\rm k}
\nonumber \\
-
&&
\alpha d^3 u_{\rm k}^{\prime \prime }
u_{\rm k}^{\prime }
+ \eta v u_{\rm k}^{\prime}
- \sin u_{\rm k} + f = 0,
\label{r2}
\end{eqnarray} 
where $c=2\pi \sqrt{g}$ is the sound speed
(in our system of units $c=d$).

Although the travelling-wave ansatz may be too crude
due to radiation of phonons by the moving kink
(e.g., see Fig.\ \ref{Fig01}a),
Eq.\ (\ref{r2}) allows to find kink's asymptotic
rigorously, because the radiation has to decay
due to nonzero damping coefficient $\eta$
in the model under consideration.
At $z\rightarrow \infty$,
substituting $u_{{\rm k}}(z) -u_f
\propto \exp (-z/d_1)$ into Eq.\ (\ref{r2}), 
we obtain for the kink width 
$d_1 (v)$ the equation 
\begin{equation}
\label{r3}
d_1^4 \cos u_f + d_1^3 \eta v 
- d_1^2 (c^2-v^2-h^2 \cos u_f)
= (h v)^2,
\end{equation}
where
$\cos u_f = \left( 1-f^2 \right)^{1/2}$. 
Similarly, we can find the tail asymptotic
behind the kink, 
$u_{{\rm k}}(z) -(u_f+2\pi p) \propto 
\exp (z/d_2)$ at 
$z \rightarrow -\infty$; 
the width $d_2 (v)$ has to satisfy the same 
equation (\ref{r3}) but with $v \rightarrow -v$.
One can see that at low velocities,
$|v| \ll c$,
the discreteness effects
lead to a decrease of the kink width 
in agreement with theory \cite{BKrev}. 
However, now Eq.\ (\ref{r3}) has a solution for
{\it any\/} kink velocity $v$. 
Thus, the discreteness effects remove 
the restriction $|v|<c$ of SG-type equations.
Now even for the classical FK model with harmonic
interaction the kink may 
move with any velocity $v$.

A kink solution corresponds
to a separatrix of the continuum limit equation.
To find the separatrix
of Eq.\ (\ref{r2}),
let us normalize the coordinate 
$\widetilde{z}=z/d$,
the velocity $\widetilde{v}=v/c$,
and define the dimensionless discreteness parameter
$\widetilde{h}=h /d
=1/ \sqrt{12 g}$. 
Introducing the new variable
$\xi =u_{{\rm k}}(\widetilde{z})$ and
the function
$w(\xi)=u_{{\rm k}}^{\prime}(\widetilde{z})$,
Eq.\ (\ref{r2}) can be rewritten as
\begin{eqnarray}
\nonumber
&&
\left\{
\widetilde{h}^2 \widetilde{v}^2 \left[ 
w^{\prime \prime \prime } (\xi)
w^{2} (\xi) +4
w^{\prime \prime } (\xi)
w^{\prime } (\xi)
w(\xi) +
\left[ w^{\prime } (\xi) \right]^{3}
\right]
\right.
\\ \nonumber
&& -
\left.
\alpha w^{\prime}(\xi) w(\xi) 
+ \left( 1-\widetilde{v}^2
-\widetilde{h}^2 \cos \xi \right) w^{\prime}(\xi)
\right.
\\
&& +
\left.
\left( \widetilde{h}^2 \sin \xi \right) w(\xi)
+ \eta \widetilde{v}
\right\} w(\xi) -\sin \xi + f = 0. 
\label{r4}
\end{eqnarray}
A (multiple) kink solution has to satisfy
the boundary conditions 
$u_{{\rm k}}(-\infty) =u_f+2\pi p$, 
$u_{{\rm k}}(+\infty) =u_f$, 
$u_{{\rm k}}^{\prime}(\pm \infty) =0$, 
or 
\begin{equation}
w(u_f+2\pi p)=w(u_f)=0.
\label{eqw1}
\end{equation}
For example, 
if $w(\xi)$ is a separatrix for the double kink
($p=2$), it has to connect the points
$(\xi=u_f+4\pi, w=0)$ and $(\xi=u_f, w=0)$.
However, due to periodicity of the substrate
potential the function $w(\xi+2\pi)$ must
correspond to the separatrix solution as well.
Thus, on the ($\xi,w$) plane the separatrices
of the multiple ($p \geq 2$) kinks must
intersect at some point with $w \neq 0$.
One can show that in the model without the discreteness effects,
$h=0$, when the phase space of Eq.\ (\ref{r4})
is two-dimensional,
such intersections are forbidden \cite{unpub}.
Thus, Eq.\ (\ref{r4}) with $h=0$
allows neither supersonic kinks
nor multiple kinks.
On the contrary, at $h \neq 0$
the phase space of Eq.\ (\ref{r4})
is four-dimensional,
thus the separatrices
corresponded to multiple kinks
may not intersect,
and multiple kinks are allowed in principle
\cite{Yu}.
Thus, {\it the discrete model formally allows
both supersonic and multiple topological excitations.}

Although we cannot find the separatrix solution
analytically, the ordinary differential equation
(\ref{r4}) can be solved numerically
with any desired accuracy.
Thus, if one could find a separatrix solution
corresponded to supersonic or multiple kink,
this will prove their existence.
Indeed, looking for a separatrix solution numerically
for the $\beta=1/\pi$ and $\eta =0.05$ case,
we found that
at small discreteness, $g=10$,
so that $\widetilde{h} \approx 0.09$
and $\alpha 
\approx 0.1$,
the separatrix solution corresponds to the
$2 \pi$-kink at forces as large as $f=0.9$,
when the kink is supersonic,
$v_{\rm k}/c \approx 1.13$.
On the other hand,
for higher discreteness,
$g=1$ so that $\widetilde{h} \approx 0.29$,
we saw the $2\pi$-kink at $f \le 0.2$
when $v_{\rm k}/c \alt 1$, and
the $4\pi$-kink at $f \ge 0.6$
when $v_{\rm k}/c > 1.3$.
{\it Thus, both supersonic kinks and multiple kinks
do exist\/}, at least for some particular
choices of model parameters.

Near kink's tails, $z\rightarrow \pm \infty $, e.g.\ for 
$\xi =u_f+2\pi n+\epsilon $, 
where $|\epsilon | \ll 1$, 
we can use the expansion 
\begin{equation}
w(\xi )=a_1\epsilon +\frac 12a_2\epsilon ^2
+\frac 16a_3\epsilon ^3+\ldots,
\end{equation}
where 
$a_1 \equiv w^{\prime }(u_f)$, 
$a_2 \equiv w^{\prime \prime }(u_f)$, 
{\it etc.\/}
Substituting this expansion into
Eq.\ (\ref{r4}) and grouping together the terms
of the same power of $\epsilon $,
we obtain the following equation for $a_1$
[cf.\ with Eq.\ (\ref{r3})],
\begin{equation}
\widetilde{h}^2
\widetilde{v}^2 a_1^4 + 
(1-\widetilde{v}^2-
\widetilde{h}^2\cos u_f) a_1^2
+\eta \widetilde{v} a_1 - \cos u_f=0,
\label{r5}
\end{equation}
which 
always has two
roots, one positive and one negative, 
for any kink velocity $v$. Then,
equating the terms for higher powers
of $\epsilon $, we obtain the
relations that uniquely determine the coefficients
$a_2$, $a_3$, {\it etc.\/}
Thus, the separatrix solution of Eq.\ (\ref{r4})
is {\it unique}, i.e.,
for a given set of system parameters
{\it the model has either the single-kink solution
or the double-kink one\/},
but never both solutions simultaneously.

Thus, we have demonstrated the existence
of supersonic and multiple kinks
in the driven discrete FK model with anharmonic interaction
(note that the discreteness of the model and
the anharmonicity of the interaction are
the necessary conditions).
But to find the parameter range for their existence
(the sufficient conditions),
we have to study numerically either the continuum-limit
equation (\ref{r4}) or, better, the original
discrete model (\ref{mot}).
The approximate variational approach
described below, essentially simplifies this task.

\section{Supersonic kinks}
\label{supersonic}

It is easy to show that
in the case of $f=\eta=h=0$,
Eq.\ (\ref{r2}) corresponds to an extremum of the following energy
functional \cite{BZKV91}, 
\begin{equation}
\label{FZ2}
E[u (z)] = \int
dz \left[ 
\frac{c^2-v^2}{2} \left( u^{\prime} \right)^2 -
\frac{\alpha d^3}{6} \left( u^{\prime} \right)^3 
-\cos u \right].
\end{equation}
Substituting a simple SG-type ansatz 
\begin{equation}
u_{\rm SG}(z) = 4 \tan ^{-1} \exp ( -z/d_{{\rm eff}})
\label{FZ4}
\end{equation}
into Eq.\ (\ref{FZ2}),
we obtain 
\begin{equation}
\label{FZ5}
E(d_{{\rm eff}}) = 4 \frac{c^2-v^2}{d_{{\rm eff}}} 
+\frac{2\pi}{3} \alpha \frac{c^3}{d_{{\rm eff}}^2} 
+4 d_{{\rm eff}}.
\end{equation}

\begin{figure}
\epsfxsize=\hsize
\epsfbox{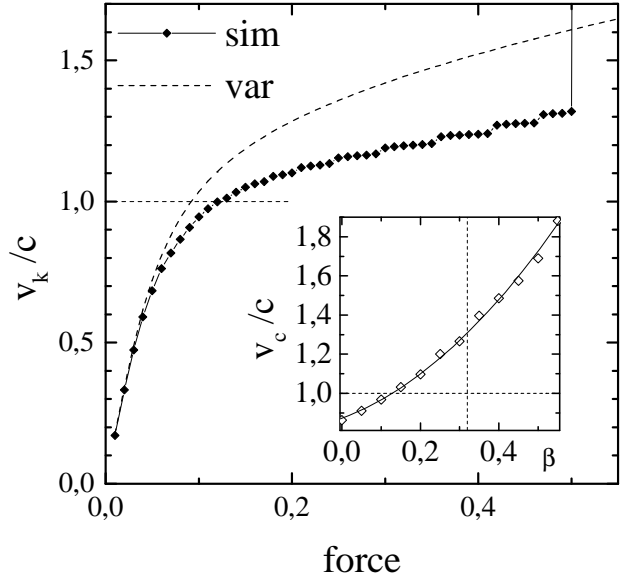}
\caption{
The velocity
$v_{\rm k}$
of the single kink versus the force 
for $\beta=1/\pi$, $g=1$, and $\eta=0.05$.
The solid curve is for simulation results, and
the dashed curve, for variational approximation.
Inset: the critical kink velocity
$v_{\rm c}(\beta)$
at the fixed force $f=0.5$.}
\label{Fig02}
\end{figure}

Although the variational approach
does not describe rigorously the kink tails
because of neglecting the discreteness effects,
it allows us to find analytically
the shape of the kink's core and,
therefore, to calculate approximately the kink velocity
for the model with anharmonic interaction.
Indeed,
looking for extrema of the function $E(d_{{\rm eff}})$,
we come to
the equation $E^{\prime }(d_{{\rm eff}}) = 0$, or 
\begin{equation}
\label{FZ6}
\kappa^3 = \left[ 1- \left( \frac{v}{c} \right)^2
\right] \kappa + \frac{\pi}{3} \alpha, 
\end{equation}
where we introduced the new variable $\kappa = d_{{\rm eff}} /d$.
For the harmonic interaction, $\alpha =0$, Eq.\ (\ref{FZ6}) has a
solution for $|v|<c$ only, which describes relativistic narrowing of the
SG kink, 
$\kappa = \left[ 1-(v/c)^2 \right]^{1/2}$.
But for the anharmonic interaction, $\alpha >0$, Eq.\ (\ref
{FZ6}) has a solution for {\it any\/} kink velocity $v$, including supersonic
velocities $|v|>c$. 
We emphasize that {\it supersonic excitations are possible for 
kinks (local compressions) only.\/}

Considering the kink as a rigid quasiparticle,
the kink effective mass can
be introduced as
(e.g., see \cite{BKrev})
\begin{equation}
m_{{\rm k}} = 
\frac{1}{a} \int_{-\infty}^{+\infty}
dz \, \left[ u^{\prime} (z) \right]^2
=\frac{4}{\pi d_{\rm eff}}.
\label{FZ7}
\end{equation}
Then, assuming that kink's
parameters at nonzero $f$ and $\eta$ are the same
as those for the $f=\eta=0$ case, 
the steady-state kink velocity can be found approximately from the
equation 
\begin{equation}
\label{FZ8a}
v_{{\rm k}} =
f/m_{{\rm k}} \eta =
\pi c f \kappa (v_{{\rm k}}) / 4 \eta.
\end{equation}
Using Eq.\ (\ref{FZ8a}),
Eq.\ (\ref{FZ6}) can be rewritten in the form
\begin{equation}
\label{FZ9}
\left[ 1 + \left( \frac{\pi f} {4 \eta } \right)^2 \right]
\kappa^3 = \kappa + \frac{\pi}{3} \alpha.
\end{equation}
Numerical solution of Eq.\ (\ref{FZ9})
allows us to find the function
$v_{\rm k}^{\rm (var)}(f)$ that is shown
by the dashed curve in Fig.\ \ref{Fig02}
together with the dependence
$v_{\rm k}(f)$
obtained by solution of the discrete
equation of motion (\ref{mot}).
One can see that in the anharmonic model
we always have
$v_{\rm k}^{\rm (var)}>c$
at $f \rightarrow 1$,
and although the simulation velocity
is lower than
$v_{\rm k}^{\rm (var)}$
due to additional damping of the moving
kink because of phonon radiation,
the discrete kink still may reach
a supersonic velocity.
Thus, the variational approach predicts that
the supersonic kinks may be expected
in the {\it anharmonic\/} FK model only.

Returning back to the {\it discrete\/} model, note that
in the classical FK model 
the driven kink cannot reach even the
sound velocity, because
it exists some critical kink velocity
$v_{{\rm c}}<c$
above which the driven kink becomes unstable
and the system goes to the ``running'' state,
where all atoms move with the velocity
$v \approx f/\eta$
\cite{BZ}.
However, 
in the anharmonic FK model, the critical kink velocity
may exceed the sound speed as
has been observed already in the simulation \cite{BBR97}.
The dependence of $v_{\rm c}$
on the anharmonicity parameter $\beta$
is shown as inset in Fig.\ \ref{Fig02}.
In this calculation we used the following algorithm
\cite{traffic}:
for a fixed value of $f$
(we took $f=0.5$),
the friction was decreased
starting from the overdamped case $\eta =1$
to the underdamped
value $\eta =10^{-3}$ in 256 steps.
At each step of $\eta$ decreasing
we waited untill the steady state was reached
and then checked if the transition to the running
state took place.

\section{Multiple kinks}
\label{multiple}

\begin{figure}
\epsfxsize=\hsize
\epsfbox{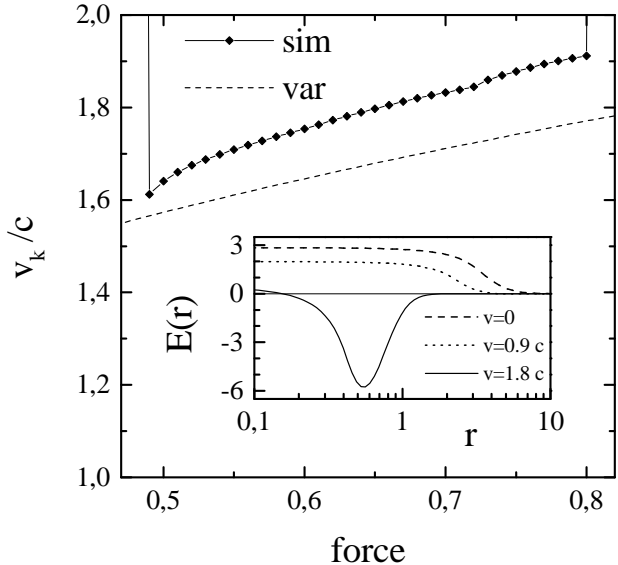}
\caption{
The same as in Fig.\ \protect\ref{Fig02}
for the double kink.
Inset: the effective energy 
$E(r)$ 
of the double kink
as function of subkink's separation $r$
for fixed kink velocities
$v_{\rm k}=0$, $0.9 \, c$, and $1.8 \, c$.
}
\label{Fig03}
\end{figure}

To study multiple kinks with the help of
a variational approach, let us
consider the double kink
as a sum of two single kinks 
separated by a distance $r$,
\begin{equation}
u_2 (z) = u_{\rm SG}(z-r/2) + u_{\rm SG}(z+r/2).
\label{dk1}
\end{equation}
Substituting the ansatz (\ref{dk1})
into the functional (\ref{FZ2}),
we obtain the effective 
energy $E(d_{{\rm eff}},r)$
which is now a function of
two parameters $d_{{\rm eff}}$ and $r$.
Looking for a minimum of $E(d_{{\rm eff}},r)$ 
over $d_{{\rm eff}}$
at $r$ fixed, we found that for
the classical FK model, $\alpha =0$,
the function 
$E(r) \equiv \min_{d_{{\rm eff}}} E(d_{{\rm eff}},r)$
is a monotonically decreasing function of $r$,
i.e., the kinks are repelled from one another.
On the other hand,
for the anharmonic FK model
the function $E(r)$ has a minimum 
at some $r=r_{\rm min} < \infty$,
so two kinks attract one another and thus
have to couple into a double kink.
The ``dissociation'' energy of the double kink
is very small at subsonic velocities, 
but becomes high enough at supersonic velocities
(see inset in Fig.\ \ref{Fig03}).
The ``size'' $r_{\rm min}$ of the double kink
decreases with $|v|$ increasing.
Thus, although the variational approach
with the SG-type ansatz is too crude,
it nevertheless predicts that
the multiple kinks could be stable
for high (supersonic) kink velocities.

As is well known, the SG kinks of the same
topological charge always repel from one another.
The same is true for static ($v=0$) kinks 
of the discrete FK model, including the anharmonic
($\alpha \ne 0$) model \cite{BKrev}
(contrary to the variational approximation
which mistakenly predicted a weak 
attraction at $|v|<c$).
Thus, it must exist a threshold kink velocity
$v_{1}$ such that at small velocities
$0 \le v < v_1$ the steady-state solution
of the model corresponds to the 
$2\pi$-kink, while at high velocities $v>v_1$,
it corresponds to the double ($4\pi$-) kink
(if $v_1$ is lower than
the kink velocity at $f=1$,
that is true at low enough values of $\eta$).
Indeed, the simulation results presented
in Fig.\ \ref{Fig03}
demonstrate that the double kink is stable within the
force interval $0.5 < f < 0.8$
but becomes unstable
at higher as well as smaller forces,
while the $2\pi$-kink is stable for
$f<0.5$ only.
Similarly one could expect the existence of
a second threshold velocity $v_2$ such that
at $v>v_2$ the steady-state solution will correspond to the 
$6\pi$-kink, {\it etc.\/}

\section{Conclusion}
\label{conclusion}

Thus, we have shown that supersonic kinks
as well as multiple kinks
do exist in the driven discrete FK model, if
the interatomic interaction is anharmonic.
Note that
both supersonic kinks and multiple kinks
remain stable at nonzero system temperatures
as well, at least for the time scale of our
numerical simulation.

Notice also that at high forces
the kink velocity is close to
that of the Toda soliton \cite{Toda}.
Indeed, the Toda soliton is characterized
by the ``jump''
$\Delta u = 2 \mu a/\beta $,
where $\mu$ is the parameter coupled with
the soliton velocity $v$ by the relationship
$v=c \sinh (\mu a) / \mu a$.
In the presence of the external substrate potential
due to boundary conditions 
the jump $\Delta u$ must be equal to $2\pi p$
for the $p$-kink; 
thus we obtain $\mu a = \pi \beta p$, or
$2 \mu =p \beta$.
In particular,
for the anharmonicity parameter $\beta =1/\pi$ 
used in the simulation, we have 
$v/c=({\rm sinh} \,p)/p \approx 1.18$
for the $2\pi$-kink and $v/c \approx 1.81$ 
for the $4\pi$-kink correspondingly.
Thus, the supersonic and multiple kinks
may be treated as Toda solitons
``disturbed'' by the external periodic potential.

\acknowledgements
Discussions with
A.\ R.\ Bishop, Yu.\ S.\ Kivshar, and M.\ Peyrard
are gratefully acknowledged.
This work was supported in part by
NATO Grant No.\ HTECH.LG 971372,
INTAS Grant No.\ 97-31061, and
by a grant from the Hong Kong Research Grants Council.


\end{document}